\documentclass[11pt]{article}
\usepackage[dvipdfmx]{graphicx}
\usepackage{color}
\begin{document}

\begin{center}
\textbf{\Large Enhancing Seasonal Adjustment Models: Constraints and Regularization for Improved Trend and AR Decomposition}

\vspace{7mm}
{\large Genshiro Kitagawa\\[3mm]
Tokyo University of Marine Science and Technology
and\\
The Institute of Statistical Mathematics
}

\vspace{3mm}
{\today}
\end{center}

\vspace{2mm}

\begin{center}{\bf\large Abstract}\end{center}

\begin{quote}
This paper investigates enhancements to model-based methods for seasonal adjustment, with a particular focus on the state space modeling framework. It addresses limitations of the standard Decomp model; specifically, the tendency to produce overly smooth trend components and the misattribution of long-term variation to the AR component when the eigenvalues of the AR model are close to unity. To mitigate these issues, the paper proposes imposing constraints on the modulus and argument of the AR eigenvalues, as well as applying regularization techniques ($L_1$ and $L_2$). These approaches are evaluated using real-world datasets. The paper is structured as follows: an overview of the Decomp model, a comparison with its noise-free variant, empirical assessment of constrained AR models, an exploration of regularization methods, and a concluding discussion of key insights.
\end{quote}

\vspace{5mm}
\noindent \textbf{Keywords:}\, Seasonal adjustment, finite trigonometric series, 
state-space model, DECOMP, AIC. \\

\section{Introduction}\label{introduction}

This paper discusses enhancements to model-based approaches for seasonal adjustment, with a particular emphasis on challenges associated with estimation in the state space modeling framework. A number of models have been proposed for seasonal adjustment (e.g., Cleveland and Tiao, 1976; Box et al., 1978; Akaike, 1979; Akaike and Ishiguro, 1982; Hillmer and Tiao, 1982; Gersch and Kitagawa, 1983; Taylor, 2010). The standard Decomp model (Kitagawa and Gersch, 1984, 1996; Kitagawa, 2021) decomposes a seasonal time series into trend, seasonal, autoregressive (AR), and observation noise components. This decomposition enables the separation of short-term random fluctuations, embedded within the seasonal structure, from the underlying trend. In particular, expressing such short-term fluctuations via an AR model while preserving a smooth trend component offers significant advantages in long-term forecasting; most notably, a substantial reduction in forecast error variance.

Despite its strengths, the Decomp model does not always yield satisfactory decompositions. In certain cases, the estimated trend becomes overly smooth, with long-term variations more appropriately attributed to the trend being absorbed into the AR component. This issue typically arises when the AR model's eigenvalues approach unity, making it difficult to distinguish between the AR and trend components. To address this limitation, we propose constraining the modulus and argument of the AR model's eigenvalues. Furthermore, we explore the application of regularization techniques ($L_1$ and $L_2$) to control the magnitude of AR parameters. These methodologies are evaluated through empirical analysis using real-world datasets.

The remainder of the paper is organized as follows. Section 2 presents an overview of the Decomp model and introduces a variant that assumes zero observation noise. Through application to two real datasets, the IIP data and the Blsallfood data, we demonstrate that the noise-free variant produces results nearly identical to the standard model. Section 3 introduces a model incorporating constraints on the modulus and argument of the AR eigenvalues, with empirical validation. Section 4 explores both $L_2$ and $L_1$ regularization approaches. Concluding remarks and a summary of key findings are provided in Section 5.

\section{Decomp Model for Seasonal Adjustment}
In the standard Decomp model, the time series with seasonality is expressed as
\begin{eqnarray}
y_n &=& T_n + S_n + p_n + w_n,
\end{eqnarray}
where $y_n$ is the time series and $T_n$, $S_n$, $p_n$ and $w_n$ are
the trend, the seasonal component, the stationary AR component and
the observation noise, respectively (Gersch and Kitagawa (1983), Kitagawa and Gersch (1984),
Kitagawa (2021)).
These components are assumed to follow the component models
\begin{eqnarray}
\Delta^{m_1}T_n &=& v_n^{(T)} \nonumber \\
\left( \sum_{j=0}^{p-1} B^j \right)^{m_2}S_n &=&  v_n^{(S)} \label{eq_component models} \\ 
p_n &=& \sum_{j=1}^{m_3} a_j p_{n-j} + v_n^{(p)},
\nonumber
\end{eqnarray}
where $B$ is the back-shift operator defined by $B x_n \equiv x_{n-1}$ and $\Delta = 1-B$
and $w_n$, $v_n^{(T)}$, $v_n^{(S)}$ and $v_n^{(p)}$ are independ Gaussian random variable
with mean 0 and the variances $\sigma^2$, $\tau_1^2$, $\tau_2^2$ and $\tau_3^2$, respectively.
Since the general form of the Decomp model includes day-of-week adjustment terms and regression components, it can be applied to various time series decomposition problems beyond seasonal adjustment. However, in this paper, we focus solely on the standard seasonal adjustment problem.

The basic observation model (1) and the component models (2) can be expressed in
a state-space model form
\begin{eqnarray}
x_n &=& Fx_{n-1} + Gv_n \nonumber \\
y_n &=& Hx_n + w_n,
\end{eqnarray}
where, for example, for $m_1=2$, $m_2=1$ and $m_3=3$, the state and the system noise are defined by 
$x_n =(T_n, T_{n-1},S_n,$ $S_{n-1},\ldots , S_{n-p+1},p_n, p_{n-1} ,p_{n-2})^T$,
and $v_n =(v_n^{(T)},v_n^{(S)},v_n^{(p)})$, and
$F$, $G$ and $H$ are defined by
{\setlength\arraycolsep{1mm}
\begin{eqnarray}
F &=& \left[ \begin{array}{cc|cccc|ccc}\arraycolsep=0.1mm
             2 & -1&   &   &   &   &   &   &   \\
             1 & 0 &   &   &   &   &   &   &   \\ \hline
               &   & -1& -1&\cdots&-1& &   &   \\
               &   & 1 &   &   &   &   &   &   \\
               &   &   &\ddots&&   &   &   &   \\
               &   &   &   & 1 &   &   &   &   \\ \hline
               &   &   &   &   &   &a_1&a_2&a_3 \\
               &   &   &   &   &   & 1 &   &\\
               &   &   &   &   &   &   & 1 &   
       \end{array}\right],\quad
G = \left[ \begin{array}{c|c|c}
             1 &   &   \\
             0 &   &   \\ \hline
               & 1 &   \\
               & 0 &   \\
               &\vdots&   \\
               & 0 &   \\ \hline
               &   & 1 \\
               &   & 0 \\
               &   & 0    
       \end{array}\right], \\
H &=& \left[ \begin{array}{cc|cccc|ccc} 
                    1 & 0 & 1 & 0 & \cdots & 0 & 1 & 0 & 0 \end{array}\right].
\end{eqnarray}
}
It is also assumed that the system noise $v_n$ and the observation noise $w_n$
are distributed according to Gaussian white noises
$v_n\sim N(0,Q)$ and $w_n \sim N(0,\sigma^2 )$, respectively.

Then, given a time series, $y_1,\ldots ,y_N$, the smoothed estimates of the components, 
$T_{n|N}$, $S_{n|N}$ and $p_{n|N}$ are obtained by the Kalman filter and the fixed-interval smoothing algorithm.
The parameters of the model $\theta = (\tau_1^2,\tau_2^2,\tau_3^2, a_1, \ldots, a_{m_3})^T$ is obtained by the maximum likelihood method and the
goodness of the model is evaluated by the AIC.
In the case of one-dimensional time series, we can run the Kalman filter by setting $R=1$
and the maximum likelihood estimate of the observation noise variance is automatically
given by (Kitagawa (2021))
\begin{eqnarray}
\hat{\sigma}^2 = \frac{1}{N}\sum_{n=1}^N (y_n - Hx_{n|n-1})^2.
\end{eqnarray}
Then the log-likelihood and the AIC are defined by 
\begin{eqnarray} 
\ell (\theta ) &=& -\frac{1}{2}\left\{ N\log 2\pi + \sum_{n=1}^N \log r_n + N \right\} \nonumber \\
{\rm AIC} &=& -2\ell (\hat{\theta}) + 2(\mbox{number of parameters of the model}),
\end{eqnarray}
where $\varepsilon_n$ and $r_n$ are one-step-ahead prediction error and 
its variance obtained by
\begin{eqnarray}
\varepsilon_n &=& y_n - H x_{n|n-1} \nonumber \\
r_n &=& \sigma^2 + H V_{n|n-1}H^T.
\end{eqnarray}
The number of parameters of the model is given by
$id(m_1)+id(m_2)+id(m_3)+1+ a_{m_3}$ where the function $id(m)$ is deifined by 
\begin{eqnarray}
id(m) = \left\{ \begin{array}{ll} 1 & {\rm if }\:\: m>0 \\
                                  0 & {\rm if }\:\: m=0 \end{array}\right. .
\end{eqnarray}

\subsection{Without Observation Noise Model}
When an AR model is expressed in state-space form, the observation noise variance is zero, i.e., $\sigma^2 = 0$. Similarly, in the context of seasonal adjustment, it is possible to assume a model without observation noise. In this subsection, we estimate both the standard Decomp model (which includes observation noise) and a noise-free variant using real-world data, and compare the resulting decompositions. As case studies, we use the IIP dataset and labor statistics data. These datasets are chosen because, under the standard Decomp model, the estimated trend tends to be overly linear or excessively smooth, leading to unsatisfactory results. They will continue to be used as representative examples in the subsequent sections.

\subsubsection{IIP Data}
We examine the monthly Index of Industrial Production of Japan from January 1979 to December 2009 (based on the 2020 standard). The number of observations is $N=$360.

\begin{table}[htbp]
\begin{center}
\caption{The log-likelihoods, $\ell(\hat\theta)$, and AIC's of the models with and without observation noise for IIP data.}
\label{Tab_IIP_AIC_with_observation_noise}

\vspace{2mm}\begin{small}
\begin{tabular}{c|cc|cc||cc|cc}
\hline
         &  \multicolumn{4}{c||}{$m_1=1$} & \multicolumn{4}{c}{$m_1=2$}  \\
\cline{2-9}
         &\multicolumn{2}{c|}{$R>0$} &\multicolumn{2}{c||}{$R=0$} &\multicolumn{2}{c|}{$R>0$} &\multicolumn{2}{c}{$R=0$} \\
   $m_3$ & $\ell(\hat\theta)$ & {\rm AIC} & $\ell(\hat\theta)$ & {\rm AIC} & $\ell(\hat\theta)$ & {\rm AIC}  & $\ell(\hat\theta)$ & {\rm AIC}  \\
\hline
   0 & 1160.15 & -2314.31 & 1142.92 & -2281.84 & 1213.64 & -2421.28 & 1202.74 & -2401.47 \\
   1 & 1184.31 & -2360.62 & 1184.31 & -2360.62 & 1215.05 & -2422.09 & 1214.36 & -2420.72 \\
   2 & 1195.90 & -2381.81 & 1187.66 & -2365.32 & 1215.76 & -2421.52 & 1215.21 & -2420.43 \\
   3 & 1204.57 & -2397.13 & 1188.53 & -2365.06 & 1225.20 & -2438.41 & 1224.91 & -2437.83 \\
   4 & 1213.48 & -2412.96 & 1199.66 & -2385.33 & 1240.02 & -2466.05 & 1240.12 & -2466.23 \\
   5 & 1213.49 & -2410.99 & 1203.32 & -2390.65 & 1242.23 & -2468.45 & 1242.40 & -2468.80 \\
   6 & 1216.29 & -2414.58 & 1221.52 & -2425.04 & 1249.50 & -2480.99 & 1249.55 & -2481.10 \\
   7 & 1224.28 & -2428.56 & 1228.23 & -2436.47 & 1251.33 & -2482.66 & 1251.68 & -2483.36 \\
   8 & 1231.28 & -2440.57 & 1235.77 & -2449.54 & 1258.83 & -2495.67 & 1258.89 & -2495.78 \\
   9 & 1252.15 & -2480.29 & 1237.61 & -2451.21 & 1259.78 & -2495.56 & 1260.64 & -2497.28 \\
  10 & 1252.86 & -2479.72 & 1239.03 & -2452.07 & 1259.85 & -2493.69 & 1262.15 & -2498.29 \\
  11 & 1256.31 & -2484.62 & 1239.03 & -2450.07 & 1260.24 & -2492.47 & 1263.99 & -2499.99 \\
  12 & 1257.52 & \textbf{-2485.05} & 1242.10 & -2454.19 & 1260.24 & -2490.47 & 1264.74 & -2499.48 \\
  13 & 1257.52 & -2483.05 & 1246.81& -2461.61 & 1263.66 & -2495.33 & 1265.03 & -2498.07 \\
  14 & 1257.52 & -2481.05 & 1246.85 & -2459.71 & 1263.87 & -2493.73 & 1272.49 & -2510.99 \\
  15 & 1257.52 & -2479.05 & 1253.60 & \textbf{-2471.19} & 1273.04 & \textbf{-2510.07} & 1274.61 & \textbf{-2513.22} \\
\hline
\end{tabular}\end{small}
\end{center}
\end{table}

Table \ref{Tab_IIP_AIC_with_observation_noise} summarizes the log-likelihood and AIC values obtained by fitting Decomp models with and without observation noise to the IIP data. The trend order ($m_1$) was set to 1 and 2, while the AR order was varied from 0 to 15. For the first-order trend model ($m_1 = 1$), the minimum AIC was attained at AR order 12 when observation noise was included ($R > 0$), and at AR order 15 when observation noise was excluded ($R = 0$). For the second-order trend model ($m_1 = 2$), the minimum AIC occurred at AR order 15 in both cases. These results suggest that the second-order trend model is preferable based on the AIC criterion.

Figure \ref{Fig_with_and_without_observation-noise} shows the decomposition results for the four cases, differentiated by the presence or absence of observation noise and the trend order ($m_1 = 1$ or $2$), with the AR order selected to minimize the AIC. The top panels correspond to the first-order trend model, while the bottom panels correspond to the second-order trend model. The left-hand panels display results with observation noise, and the right-hand panels without it.

For the first-order trend model ($m_1 = 1$), the trend component reduces to a constant, regardless of whether observation noise is present. Consequently, the AR component absorbs most of the data's long-term trend. In contrast, for the second-order trend model ($m_1 = 2$), the trend component captures the general structure of the time series, although multi-year cyclical fluctuations are absorbed into the AR component rather than the trend.

In the first-order trend case, the trends and seasonal components in the models with and without observation noise are nearly identical. Furthermore, the AR component in the noise-free model appears to approximate the sum of the AR and observation noise components in the model with noise. In the second-order trend case, the estimated observation noise is negligible, and the decomposition results are virtually identical. These findings indicate that, at least for this dataset, models with and without observation noise yield almost equivalent decompositions.

\begin{figure}[tbp]
\begin{center}
\includegraphics[width=140mm,angle=0,clip=]{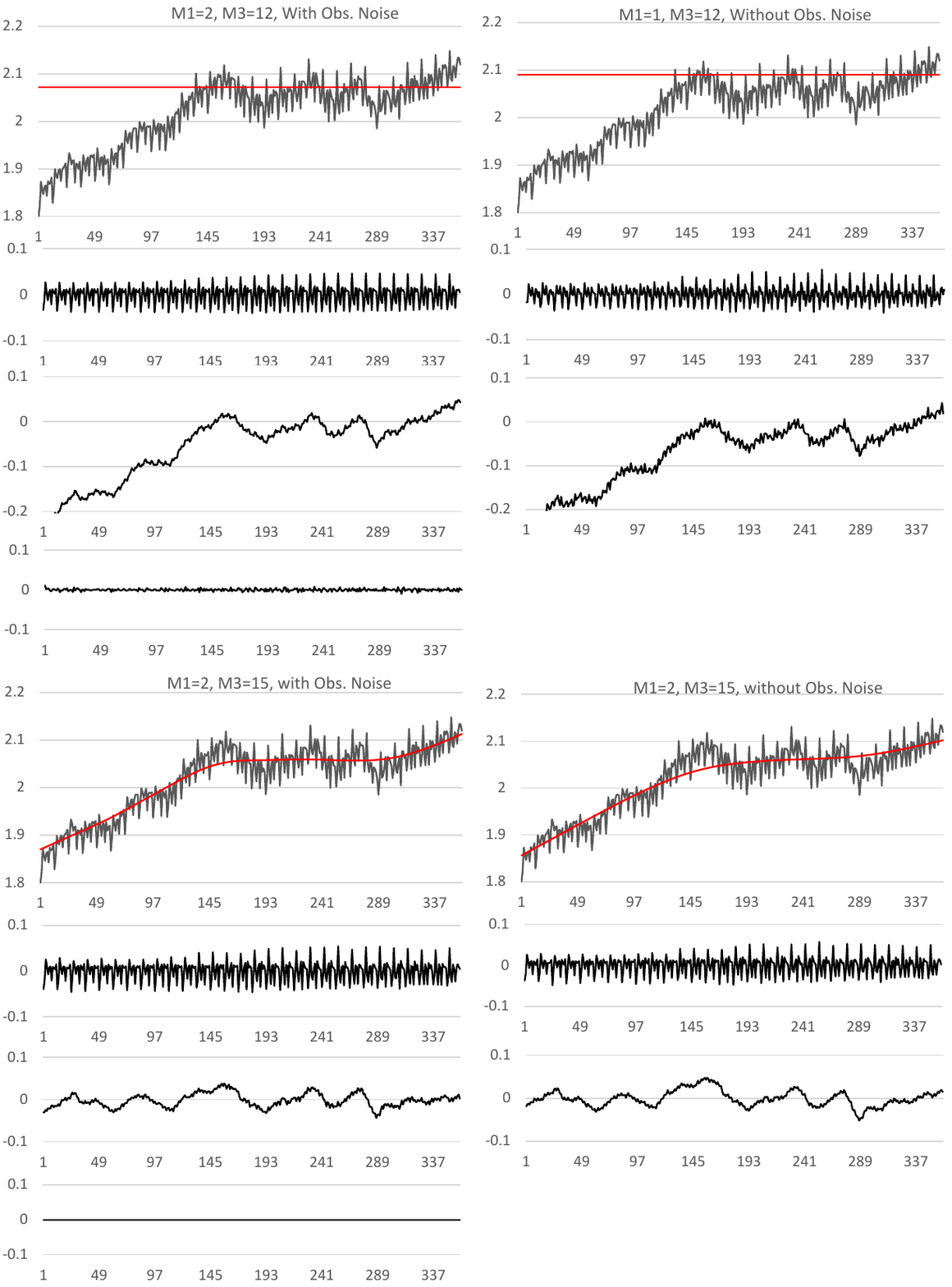}
\caption{Comparison of seasonal adjustment with observation noise (left) and without observation noise (right) for IIP data. Upper plots: $m_1=1$, bottom plots: $m_1=2$. In each plots, the panels from top to bottom show the original data and the trend component, the seasonal component, the AR component, and the observation noise component, respectively.}\label{Fig_with_and_without_observation-noise}
\end{center}
\end{figure}

\newpage
\subsubsection{Blsallfood Data}
Blsallfood data is the number of employees in food industries from January 1967 to December 1979 released by the US Bureau of Labor Statistics.

\begin{table}[htbp]
\begin{center}
\caption{The log-likelifoods, $\ell(\hat\theta)$, and AIC's of the models with and without observation noise for Blsallfood data. $m_1$ is the order of the trend model.}
\label{Tab_blsfood_AIC_with_observation_noise}

\vspace{2mm}\begin{small}
\begin{tabular}{c|cc|cc||cc|cc}
\hline
         &  \multicolumn{4}{c||}{$m_1=1$} & \multicolumn{4}{c}{$m_1=2$}  \\
\cline{2-9}
         &\multicolumn{2}{c|}{$R>0$} &\multicolumn{2}{c||}{$R=0$} &\multicolumn{2}{c|}{$R>0$} &\multicolumn{2}{c}{$R=0$} \\
   $m_3$ & $\ell(\hat\theta)$ & {\rm AIC} & $\ell(\hat\theta)$ & {\rm AIC} & $\ell(\hat\theta)$ & {\rm AIC}  & $\ell(\hat\theta)$ & {\rm AIC}  \\
\hline
   0 & -660.55 & 1325.10 &-661.06 & 1326.12 & -636.89 & 1277.78 & -637.03 & 1278.05 \\
   1 & -614.95 & 1237.90 &-614.77 & 1237.54 & -582.43 & \textbf{1172.86} & -582.49 & \textbf{1172.99} \\
   2 & -610.97 & 1231.93 &-610.75 & 1231.51 & -582.28 & 1174.57 & -582.27 & 1174.54 \\
   3 & -610.87 & 1233.74 &-610.65 & 1233.30 & -582.10 & 1176.19 & -582.09 & 1176.18 \\
   4 & -610.27 & 1234.54 &-610.07 & 1234.14 & -582.03 & 1178.06 & -582.03 & 1178.05 \\
   5 & -610.01 & 1236.02 &-609.81 & 1235.62 & -581.78 & 1179.55 & -581.76 & 1179.51 \\
   6 & -605.85 & 1229.70 &-606.02 & 1230.04 & -581.46 & 1180.92 & -581.42 & 1180.85 \\
   7 & -604.57 & 1229.13 &-604.72 & 1229.44 & -580.65 & 1181.30 & -580.62 & 1181.23 \\
   8 & -601.40 & \textbf{1224.80} &-601.56 & \textbf{1225.13} & -580.35 & 1182.70 & -580.35 & 1182.69 \\
   9 & -601.39 & 1226.78 &-601.55 & 1227.11 & -578.35 & 1180.70 & -578.29 & 1180.58 \\
  10 & -601.39 & 1228.77 &-601.54 & 1229.09 & -577.80 & 1181.61 & -577.79 & 1181.58 \\
\hline
\end{tabular}\end{small}
\end{center}
\end{table}

Table \ref{Tab_blsfood_AIC_with_observation_noise} presents the results of fitting Decomp models with and without observation noise to the Blsallfood data. Trend orders of 1 and 2, along with AR orders ranging from 0 to 10, were evaluated. For this dataset, the AIC values are nearly identical regardless of the inclusion of observation noise. When the trend order is set to 1, the minimum AIC occurs at AR order 8, while for a trend order of 2, the minimum AIC is achieved at AR order 1. Comparing the two trend specifications, the second-order trend model is clearly preferred based on the AIC criterion.

\begin{figure}[tbp]
\begin{center}
\includegraphics[width=140mm,angle=0,clip=]{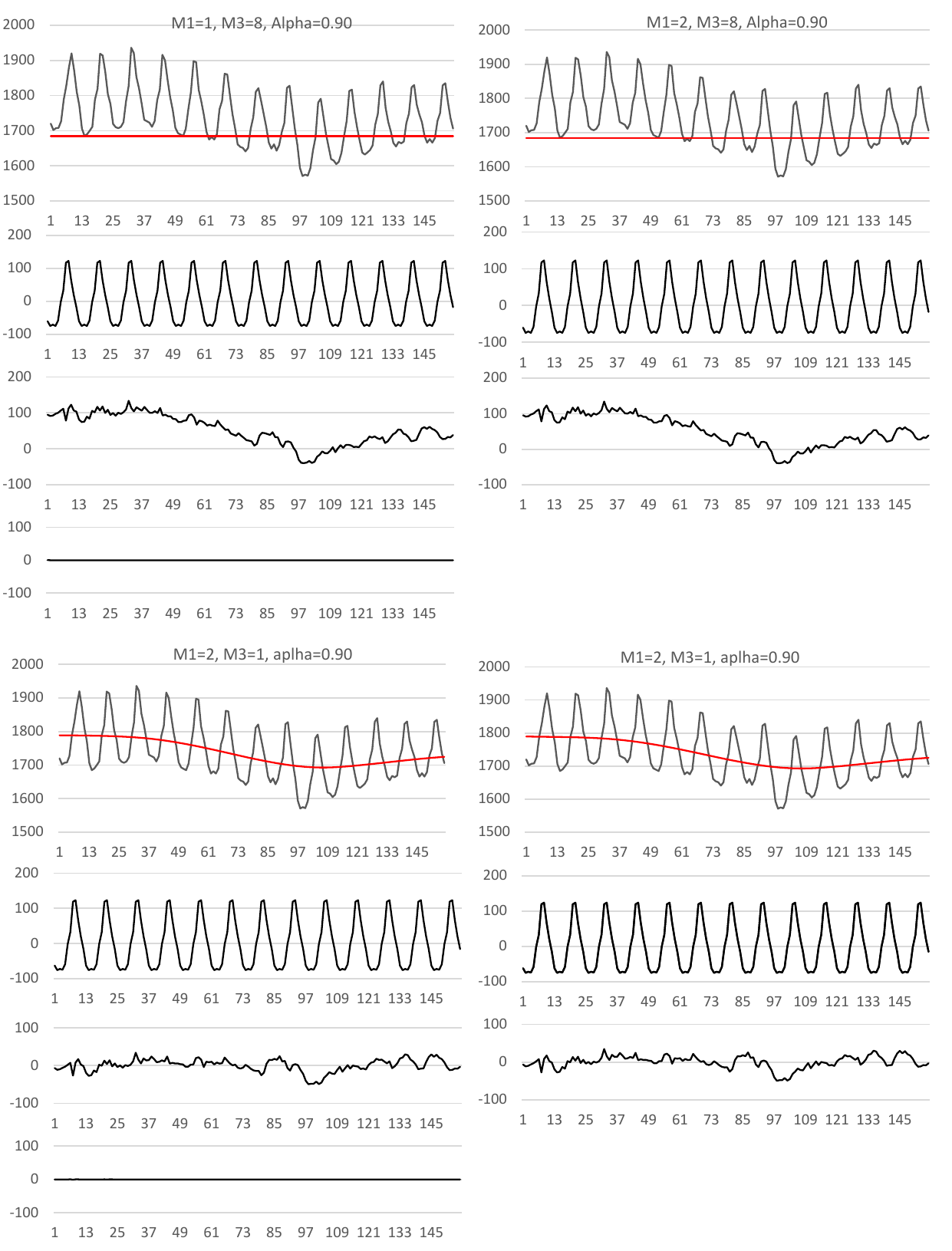}
\caption{Comparison of seasonal adjustment with observation (left) noise and without observation (right) for Blsallfood data. Upper plots: $m_1=1$, right: $m_1=2$. }\label{Fig_with_and_without_observation-noise}
\end{center}
\end{figure}
Figure \ref{Fig_with_and_without_observation-noise} presents the decomposition results from the seasonal adjustment model, where the AR order corresponds to the minimum AIC. The results indicate that models with and without observation noise yield nearly identical decompositions. When the trend order is set to 1, the trend component reduces to a constant, and the trend information is largely absorbed into the AR component. In contrast, with a trend order of 2, the trend captures the long-term structure of the time series, while the short-term fluctuations are represented by the AR component.

These examples demonstrate that, when an AR component is included, assuming zero observation noise leads to decomposition results that are practically equivalent to those obtained with observation noise. Accordingly, in the subsequent sections, we consider only models without observation noise.

Furthermore, since the seasonal component is generally unaffected by variations in the trend and AR components, it will be omitted from the figures in the remaining examples.

\newpage
\section{AR Model with Restricted Phase and Amplitude}

The AR model is highly flexible, and as the absolute values of its characteristic roots approach 1, it becomes increasingly difficult to distinguish the AR component from the trend component. This explains why, when using a first-order trend model, the estimated trend often reduces to a constant, and with a second-order trend model, the trend may become excessively smooth.
To address this issue, the standard Decomp model typically imposes a constraint that the absolute values of the PARCOR coefficients must remain below a specified threshold, such as 0.9.
However, as illustrated by the examples in the previous section, this constraint does not always produce satisfactory results.
In this section, we propose an alternative approach that directly constrains both the modulus and the phase angles of the AR model's characteristic roots.

Given an AR component model
\begin{eqnarray}
p_n = \sum_{k=1}^{m_3} a_j p_{n-j} + v_n^{(p)},
\end{eqnarray}
the characteristic roots $\lambda_j, \,j=1,\ldots,m_3$ are defined as the solutions of the characteristic equation
\begin{eqnarray}
a(\lambda) = \lambda^{m_3} - \sum_{j=1}^{m_3} a_j \lambda^{m_3-j} = 0.
\end{eqnarray}
In general, the characteristic roots consist of $m_r$ real roots and $m_i$ complex conjugate pairs, such that $m_3 = m_r + 2m_i$.
We express each complex root as
$ \alpha_j + i\beta_j = r_j e^{\theta_j}$, where $r_j$ is the modulus (or absolute value) and $\theta_j$ is the argument (or phase angle).
The AR model is stationary if all its characteristic roots lie strictly within the unit circle in the complex plane, i.e., $r_j < 1$ for all $j$.
However, even when the stationarity condition is satisfied, if the characteristic roots are close to the unit circle (i.e., $r_j \equiv 1$), the decomposition of the time series becomes ambiguous. This is because the AR component may closely resemble a trend component, making it difficult to distinguish between the two.

\begin{figure}[h]
\begin{center}
\includegraphics[width=70mm,angle=0,clip=]{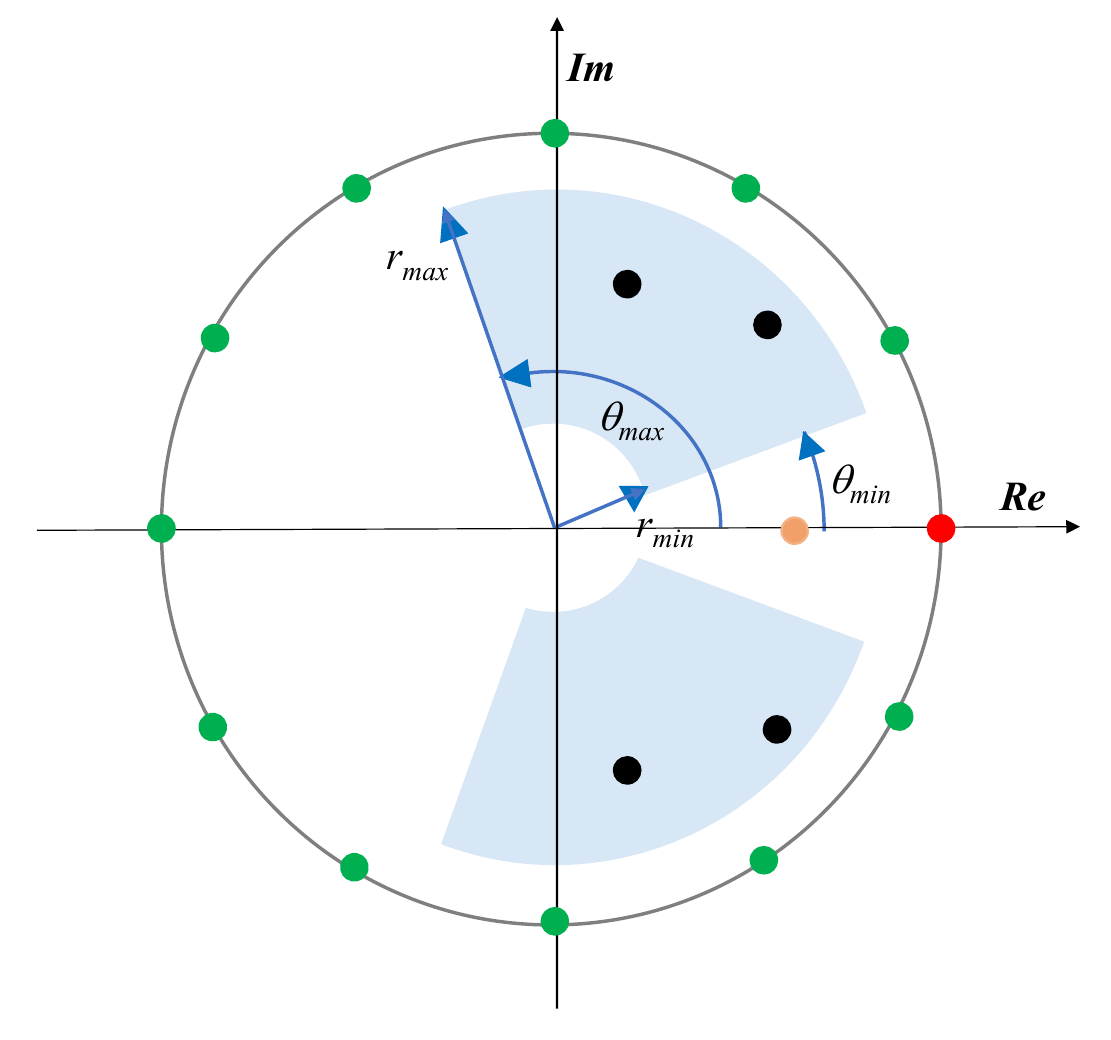}
\includegraphics[width=40mm,angle=0,clip=]{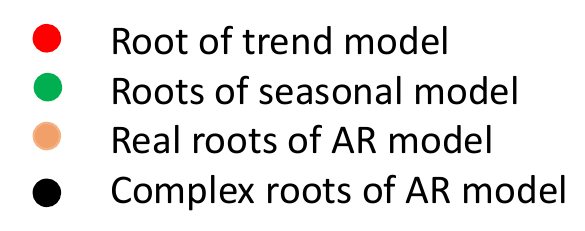}
\end{center}
\caption{Restriction of argument (phase angle) and modulus (absolute value) of the characteristic roots of AR model. }\label{Fig_Restricted_AR_model}
\end{figure}

To mitigate this issue, we impose constraints on both the modulus and the phase angle of the characteristic roots of the AR component model, as follows:
\begin{eqnarray}
  r_{min} < r_j < r_{max}  \nonumber \\
  \theta_{min} < \theta_j < \theta_{max} .
\end{eqnarray}
Figure \ref{Fig_Restricted_AR_model} illustrates this constraint region, where the black dots represent a case in which four characteristic roots lie within the specified bounds. The pink dot represents an example of a real root that satisfies the modulus constraint.

Given $m_i$ pairs of complex roots, $r_j$ and $\pm\theta_j$, $j=1,\ldots,m_i$, and $m_r$ real roots satisfying 
$r_{min} < s_j < r_{max}$, $j=1,\ldots,m_r$, the coefficients of the autoregressive model, $a_j$,  are obtained by
\begin{eqnarray}
  a(\lambda) &=& \prod_{j=1}^{m_r} (\lambda -s_j) \prod_{j=1}^{m_i} (\lambda^2 - 2r_j \cos(\theta_j)\lambda + r^2) \nonumber \\
     &\equiv& \lambda^{m_3} - \sum_{j=1}^{m_3} a_j \lambda^{m_3-j}  .
\end{eqnarray}

\newpage
\subsection{Example:  IIP data}
Table \ref{Tab_AIC_IIP} presents the results of applying the Decomp model without observation noise (denoted as AR$\_$type = 1), as introduced in the previous section, and a version with constraints on the characteristic roots of the AR model (AR$\_$type = 2) to the IIP data. Two trend orders are considered: $m_1 = 1$ and $m_1 = 2$.

For $m_1 = 1$, the minimum AIC for AR$\_$type = 1 is achieved at AR order 9, while for AR$\_$type = 2, the minimum AIC occurs at $m_r = 1$ and $m_i = 3$, yielding $m_3 = 1 + 2 \times 3 = 7$.
For $m_1 = 2$, the minimum AIC for AR$\_$type = 1 is again achieved at AR order 9. For AR$\_$type = 2, the lowest AIC is obtained when $m_r = 2$ and $m_i = 2$, resulting in $m_3 = 6$. However, the AIC values for the configurations $(m_r = 0,, m_i = 2)$, $(m_r = 1,, m_i = 2)$, and $(m_r = 2,, m_i = 2)$ are nearly indistinguishable, indicating that the models offer comparable goodness of fit.

Given that AR$\_$type = 2 imposes additional constraints on the AR model coefficients, it is reasonable that its AIC values are generally higher than those of AR$\_$type = 1.

\begin{table}[htbp]
\begin{center}
\caption{The AIC's of the AR$\_$type=1 and AR$\_$type=2 models with various orders for IIP data}
\label{Tab_AIC_IIP}

\vspace{2mm}
\begin{small}\tabcolsep=1mm
\begin{tabular}{c|c|ccc||c|ccc}
      &  \multicolumn{4}{|c||}{$m_1=1$} & \multicolumn{4}{|c}{$m_1=2$} \\  \cline{2-9}
$m_3$ &    & \multicolumn{3}{|c||}{AR$\_$type=2} &   & \multicolumn{3}{|c}{AR$\_$type=2} \\  \cline{2-9}
      & $m_1=1$ & $m_r=0$ & $m_r=1$ & $m_r=2$ & $m_1=2$ & $m_r=0$ & $m_r=1$ & $m_r=2$  \\
\hline
0 &	-2281.84 & -2281.84 &		   &          & -2401.47 & -2401.47 &		   &	       \\
1 & -2359.86 &          & -2343.01 &		  & -2420.72 &          & -2415.04 & 		   \\
2 & -2365.41 & -2326.03 & & -2335.83 & -2420.43 & -2412.04 &		   & -2403.52  \\
3 & -2364.03 &          & -2339.72 & 		  & -2437.83 &          & -2434.84 &		   \\
4 & -2384.03 & -2333.66 &		   & -2322.33 & -2466.23 & \textbf{-2439.61} &		   & -2430.36  \\
5 &	-2389.89 &          & -2349.41 &          & -2468.80 &          & \textbf{-2439.35} &  \\
6 & -2423.87 & -2329.27 &		   & -2332.27 & -2481.10 & -2435.53 & & \textbf{-2439.65}  \\
7 & -2434.89 &      &\textbf{-2350.13}&		  & -2483.36 &         & -2438.03 &		   \\
8 & -2448.50 & -2326.96 & &	-2343.46& -2495.78 & -2434.17 &          &	-2438.46 \\
9 & \textbf{-2451.30} &   & -2347.15 & 		  & \textbf{-2497.28} &          &          &           \\
\hline \end{tabular}
\end{small}\end{center}
\end{table}

\begin{figure}[tbp]
\begin{center}
\includegraphics[width=140mm,angle=0,clip=]{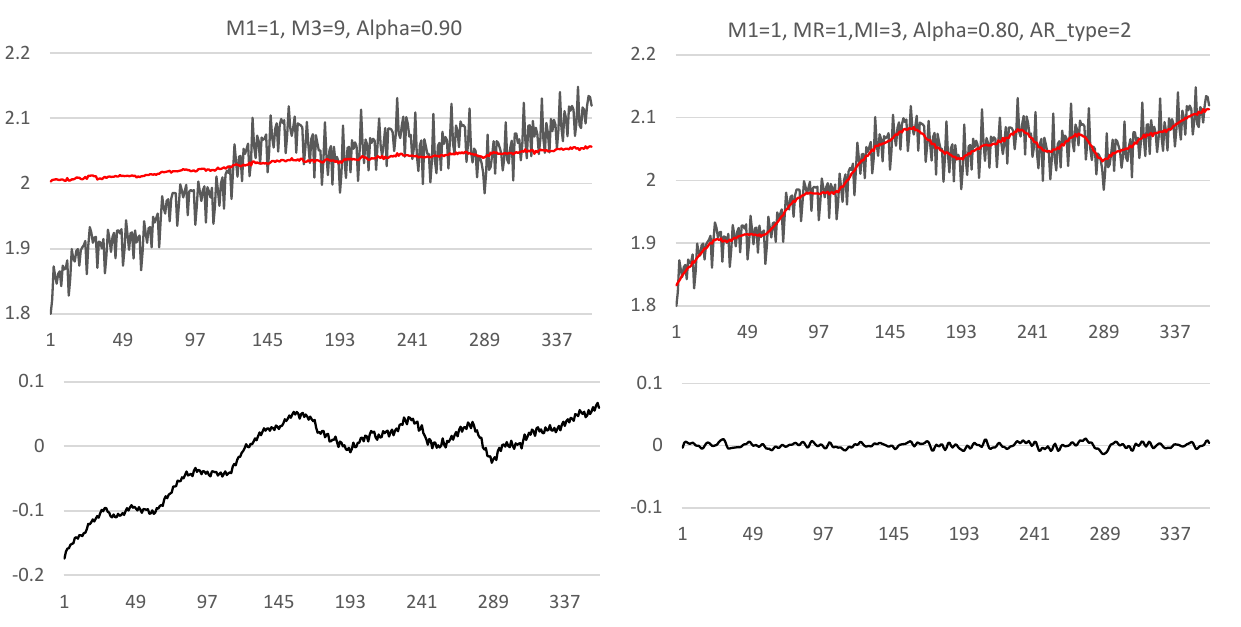}
\includegraphics[width=140mm,angle=0,clip=]{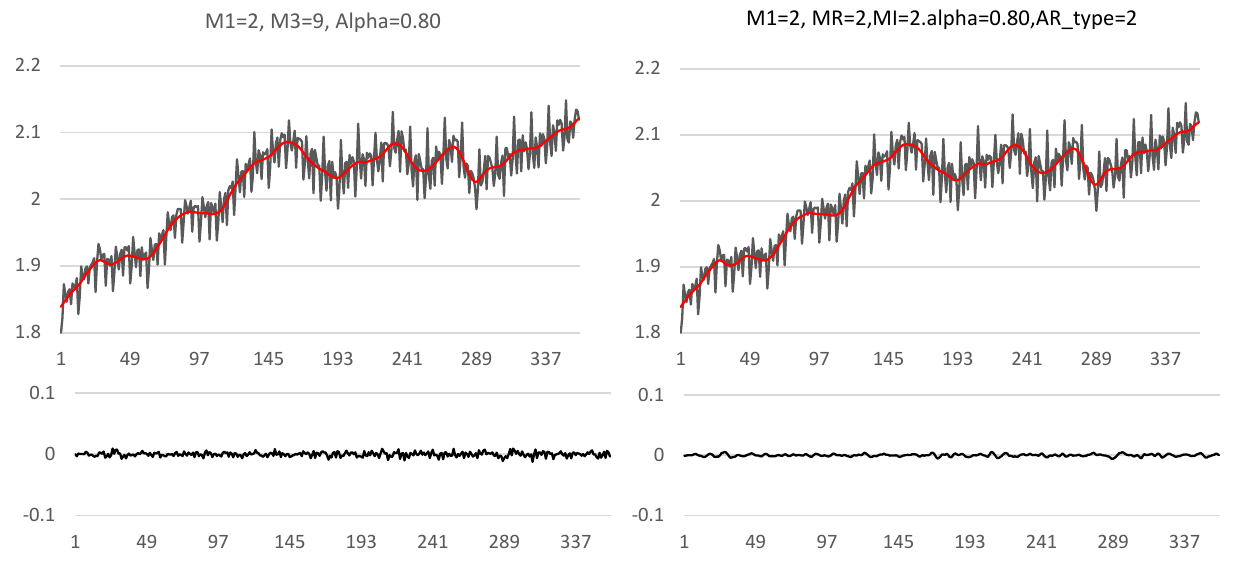}
\end{center}
\caption{Decomposition of IIP data by AR$\_$type=1 model (left plots) and AR$\_$type=2 model (right plots). Upper plots: first order trend, $m_1=2$, right plots: second order trend, $m_1=2$. }\label{Fig_compare AR-type=1 and 2}
\end{figure}

Figure \ref{Fig_compare AR-type=1 and 2} shows the decomposition results obtained using the AR$\_$type = 1 and AR$\_$type = 2 models. The top panels correspond to the first-order trend case ($m_1 = 1$), while the bottom panels represent the second-order trend case ($m_1 = 2$). The left-hand panels depict the results for AR$\_$type = 1, and the right-hand panels for AR$\_$type = 2.

When the trend order is 1 ($m_1 = 1$), the trend component in the AR$\_$type = 1 model is almost linear. In contrast, under AR$\_$type = 2 with $m_r = 1$ and $m_i = 3$, the variation in the AR component is reduced, allowing the model to better capture the underlying trend of the IIP data. Long-period variations that were previously absorbed into the AR component are now more appropriately reflected in the trend component.

When the trend order is 2 ($m_1 = 2$), the AR$\_$type = 1 model produces a smooth trend that reasonably captures the general tendency of the time series. However, the corresponding AR component shows relatively large fluctuations. In the AR$\_$type = 2 model with $m_r = 2$ and $m_i = 2$, the trend component closely resembles that of the $m_1 = 1$ case and effectively expresses the overall pattern of the IIP data. Since the trend components are nearly identical in both models, the reduction in AR component variability in AR$\_$type = 2 suggests that some of this variation has been absorbed by the seasonal component.

\subsection{Example: Blsallfood data}

Table \ref{Tab_AIC_Blsallfood} presents the results of applying the Decomp model without observation noise (AR$\_$type = 1), as described in the previous section, and the model with constraints on the characteristic roots of the AR model (AR$\_$type = 2) to the BLSallfood data.
For the first-order trend model ($m_1 = 1$), the minimum AIC under AR$\_$type = 1 is obtained at AR order 8, while under AR$\_$type = 2, the optimal configuration is $m_r = 1$ and $m_i = 1$, corresponding to $m_3 = 1 + 2 \times 1 = 3$.
For the second-order trend model ($m_1 = 2$), the minimum AIC under AR$\_$type = 1 occurs at AR order 1, whereas under AR$\_$type = 2, the best result is achieved with $m_r = 1$ and $m_i = 0$, giving $m_3 = 1$.

\begin{table}[htbp]
\begin{center}
\caption{The AIC's of the AR$\_$type=1 and AR$\_$type=2 models with various orders for Blsallfood data. $m_1$ is the order of the trend model.}
\label{Tab_AIC_Blsallfood}

\vspace{2mm}
\begin{small}
\tabcolsep=1mm
\begin{tabular}{c|c|ccc||c|ccc}
      &  \multicolumn{4}{|c||}{$m_1=1$} & \multicolumn{4}{|c}{$m_1=2$} \\  \cline{2-9}
$m_3$ &    & \multicolumn{3}{|c||}{AR$\_$type=2} &    &\multicolumn{3}{|c}{AR$\_$type=2} \\  \cline{2-9}
      & $m_1=1$ & $m_r=0$ & $m_r=1$ & $m_r=2$ & $m_1=2$ & $m_r=0$ & $m_r=1$ & $m_r=2$  \\
\hline
0 &	1326.12 & 1328.12 &		    &         & 1278.05 & 1280.05 &		   &	       \\
1 & 1237.54 &         & 1258.65 &         & \textbf{1172.99}  &          & \textbf{1174.56} & 		   \\
2 & 1231.51 & 1256.87 &		    & 1296.75 & 1174.54 & 1188.58 &		   & 1186.57  \\
3 & 1233.30 &      &\textbf{1252.29} &    & 1176.18 &          & 1184.52 &		   \\
4 & 1234.14 & 1259.91 &		    & 1260.78 & 1178.05 & 1179.96 &		   & 1179.33  \\
5 &	1235.62 &         & 1255.57 &		  & 1179.51 &          & 1180.47 &		   \\
6 & 1230.04 & 1265.22 &		    & 1262.96 & 1180.85 & 1182.58 & 		   & 1185.38  \\
7 & 1229.44 &         & 1258.49 &		  & 1181.23 &        & 1184.04 &		   \\
8 & \textbf{1225.13}& 1265.37 & & 1263.83 &	1182.69 & 1186.40 &          & 1186.18\\
9 & 1227.11 &         & 1262.37 & 		  & 1180.58 &          &  1187.87 &           \\	
10& 1229.09 & 1268.71 &        & 1266.45 & 1181.58 & 1190.36 &      & 1189.89 \\
		\hline \end{tabular}
\end{small}
\end{center}
\end{table}

\begin{figure}[tbp]
\begin{center}
 \includegraphics[width=140mm,angle=0,clip=]{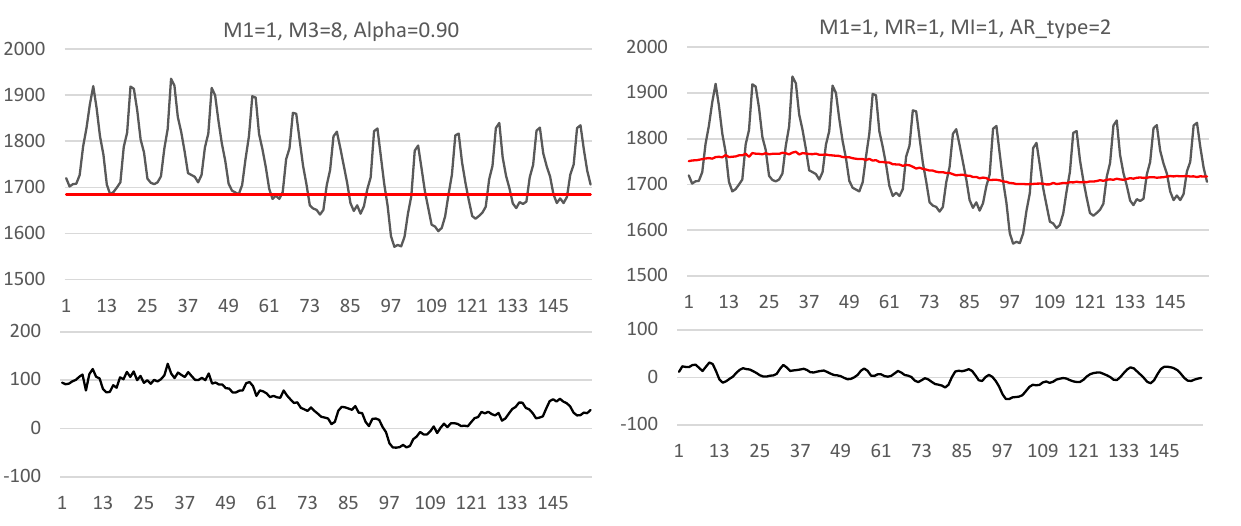}
 \includegraphics[width=140mm,angle=0,clip=]{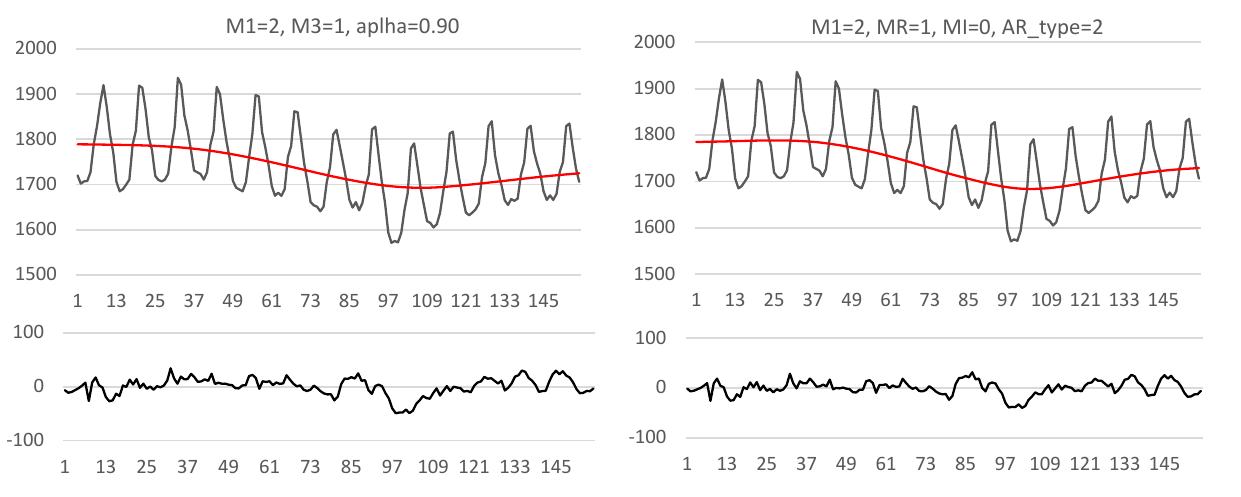}
\caption{Decomposition of Blsallfood data by AR$\_$type=1 model (left plots) and AR$\_$type=2 model (right plots). Upper plots: first order trend model, $m_1=2$, 
bottom plots: sesond order trend model, $m_1=2$. }\label{Fig_blsfood_AR-type=2}
\end{center}
\end{figure}

Figure \ref{Fig_blsfood_AR-type=2} compares the decomposition results obtained using the AR$\_$type = 1 and AR$\_$type = 2 models. The upper panels correspond to the case of a first-order trend ($m_1 = 1$), and the lower panels to a second-order trend ($m_1 = 2$).

When $m_1 = 1$, the trend estimated by the AR$\_$type = 1 model is essentially constant, with long-period fluctuations absorbed into the AR component. In contrast, the AR$\_$type = 2 model yields a trend that better follows the overall trajectory of the time series, resulting in a noticeable reduction in long-period components within the AR term. However, this trend still appears somewhat overly smooth and fails to fully capture the finer fluctuations present in the data.

For $m_1 = 2$, the trend estimated by the AR$\_$type = 1 model already reflects the general tendency of the time series, comparable to the $m_1 = 1$ case under the AR$\_$type = 2 model. Nonetheless, the trend obtained with the AR$\_$type = 2 model more effectively captures both the long-term and intermediate variations, offering a better representation of the underlying data dynamics.

The findings of this section offer several key insights. In conventional Decomp models, as demonstrated through the two datasets analyzed in this study, the decomposition into trend and AR components is often highly sensitive to the choice of AR order. In particular, selecting the AR order based on the minimum AIC can lead to excessively smooth trend components, suggesting that the decomposition becomes unstable as higher-order AR models are introduced. However, as shown in this paper, the AR$\_$type = 2 model where constraints are imposed on the eigenvalues of the AR component provides a more stable and interpretable decomposition, even under such conditions.

\newpage
\section{Regularization Method}
In the regularization approach, a penalty term $R(\theta)$ is added to the log-likelihood function to control model complexity. Specifically, model parameters are estimated by minimizing the following objective:
\begin{eqnarray}
\log \ell(\theta)+\lambda R(\theta),
\end{eqnarray}
where $\ell(\theta)$ is the likelihood function of the model parameterized by $\theta$, and
$\lambda$  is a tuning parameter that determines the strength of the regularization. 
The regularization term $R(\theta)$ is typically designed to penalize model complexity and prevent overfitting. In what follows, we consider both $L_2$ and $L_1$ regularization.

\subsection{AR Modeling with $L_2$ Regularization Term}
In this subsection, we focus on the $L_2$ regularization approach, where parameters are estimated by minimizing
\begin{eqnarray}
  \log\ell (\theta) + \lambda R_2(\theta).
\end{eqnarray}
Although a natural choice for the penalty term is $R_2(\theta) = \sum_{j=1}^{n_v+m_3} \theta_{j}^2 $, we instead adopt
\begin{eqnarray}
  R_2(\theta) =\sum_{j=1}^{m_3} \theta_{n_v+j}^2  , 
\end{eqnarray}
which specifically targets the autoregressive (AR) coefficients.
In time series modeling, it may be more conventional to penalize the AR coefficients $\sum_{j=1}^{m_3} a_{j}^2$ or the partial autocorrelation (PARCOR) coefficients  $\sum_{j=1}^{m_3} b_{j}^2$.  However, empirical findings suggest that directly penalizing the relevant subset of parameters, as in the formulation above, tends to simplify the optimization process involved in parameter estimation.

\begin{figure}[htbp]
\begin{center}
\includegraphics[width=140mm,angle=0,clip=]{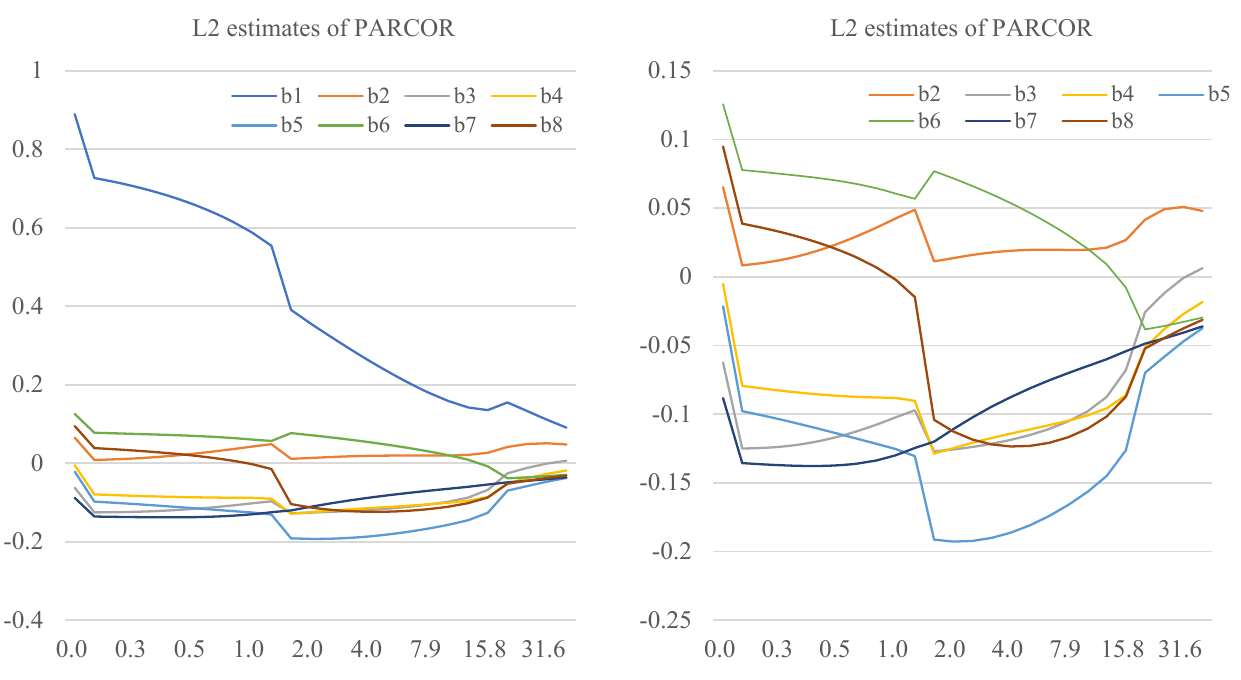}
\end{center}
\caption{Change of estimated parameters under $L_2$ regularization. $\lambda = 0$, and $10^{j/10},
j=-8,\ldots ,16$ }\label{Fig_L2-estimates}
\end{figure}

The left panel of Figure \ref{Fig_L2-estimates} illustrates how the partial autocorrelation (PARCOR) coefficients vary with respect to the regularization weight parameter $\lambda$ in the model with a second-order trend ($m_1=2$)  and AR order $m_3=8$, applied to the Blsallfood data. The horizontal axis represents the value of $\lambda$, while the vertical axis shows the corresponding PARCOR coefficients. As $\lambda$ increases, the PARCOR values exhibit a smooth and gradual decline. The right panel provides a closer examination by displaying the trajectories of seven PARCOR coefficients, excluding $b_1$, which has the largest absolute value.

\begin{figure}[tbp]
\begin{center}
\includegraphics[width=140mm,angle=0,clip=]{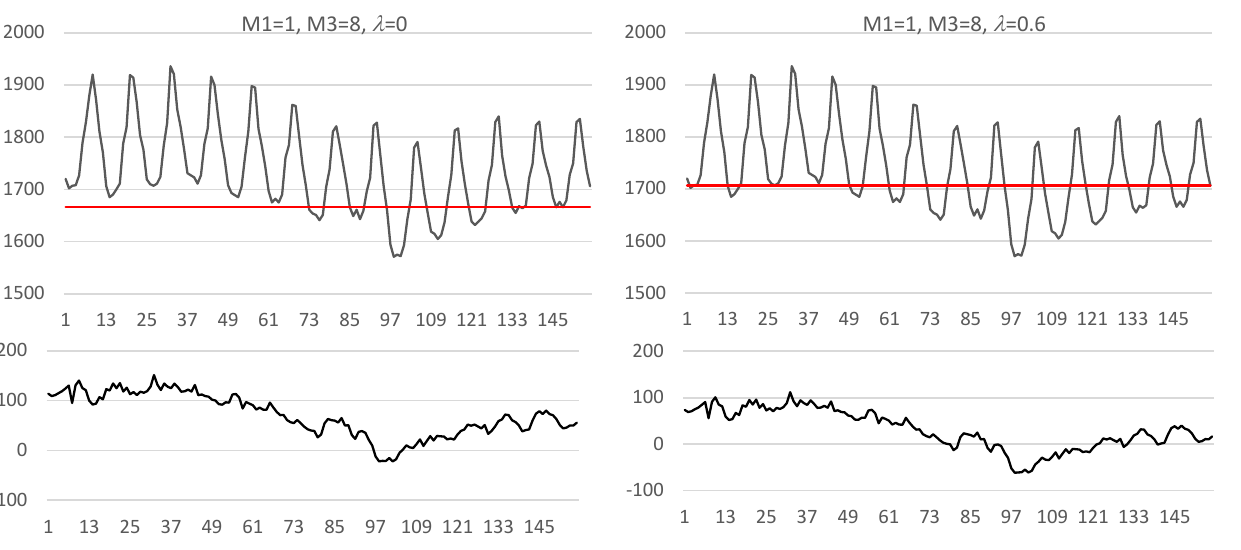}
\includegraphics[width=140mm,angle=0,clip=]{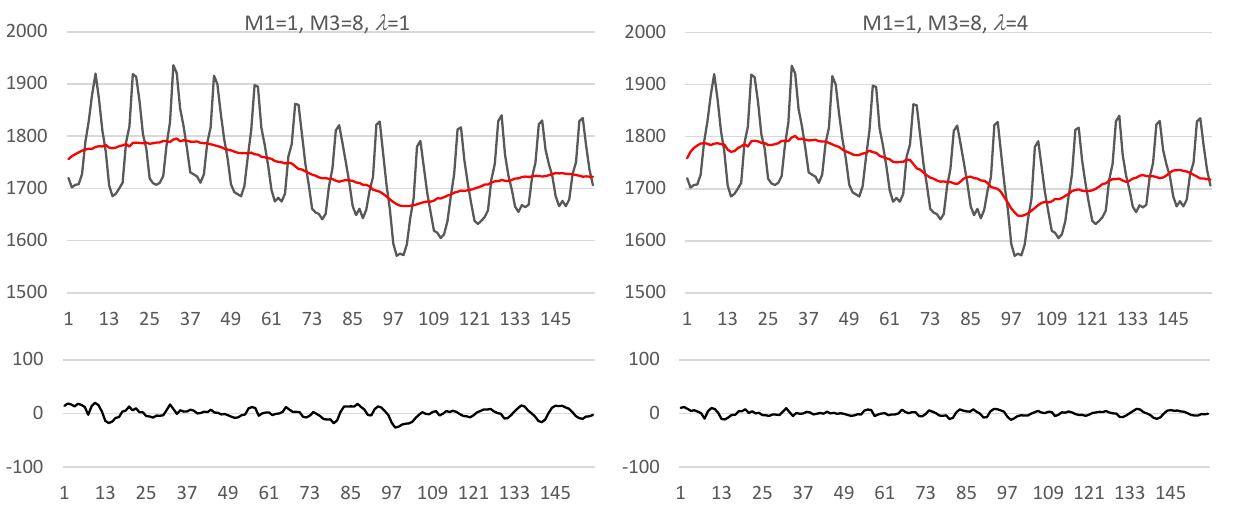}
\end{center}
\caption{Decomposition of time series by first order trend under various $L_2$ regularization terms. $\lambda = 0$ 0.6, 1 and 4 }\label{Fig_L2-estimates_m1=1}
\end{figure}

Figure \ref{Fig_L2-estimates_m1=1} presents the trend and AR components for the case of $m_1=1$
and $m_3=8$ under four different values of the regularization parameter $\lambda$=0, 0.6, 1 and 4. When $\lambda$=0.6, the trend remains nearly constant, resembling the case without regularization ($\lambda=0$), and the AR component still includes long-period fluctuations. In contrast, at 
$\lambda=1$, the trend more effectively captures the overall pattern of the data, while the AR component shows a reduction in the long-period fluctuations observed in the unregularized case. When $\lambda=4$, the trend exhibits slightly greater variation, while the AR component becomes noticeably smaller.

\begin{figure}[tbp]
\begin{center}
\includegraphics[width=140mm,angle=0,clip=]{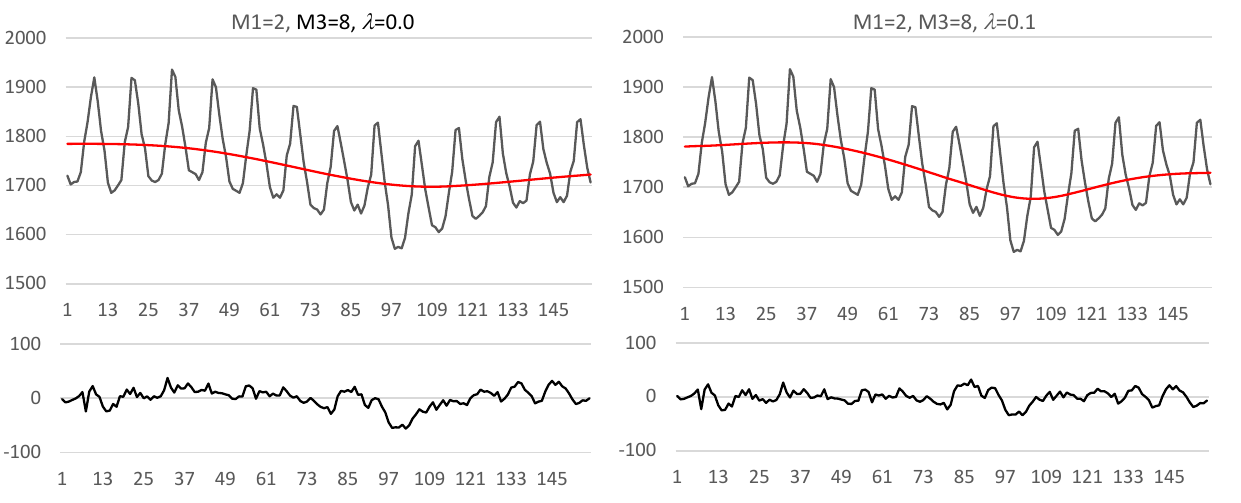}
\includegraphics[width=140mm,angle=0,clip=]{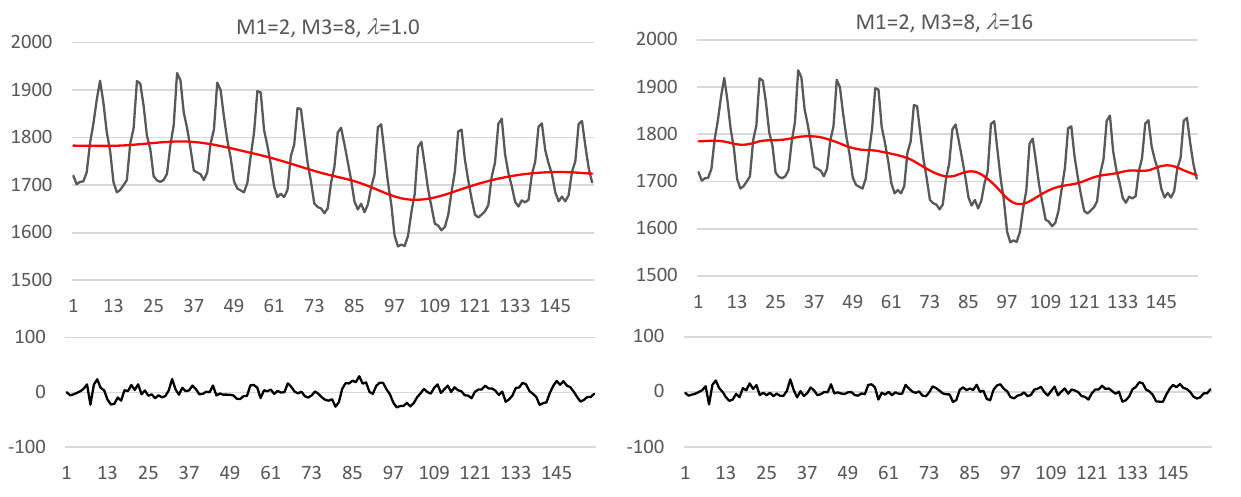}
\end{center}
\caption{Decomposition of time series by second order trend under various $L_2$ regularization terms. $\lambda = 0$ 0.1, 1 and 16}\label{Fig_L2-estimates_M1=2}
\end{figure}

Figure \ref{Fig_L2-estimates_M1=2} presents the results for the case of the second order trend, $m_1=2$, with the AR model order set to $m_3=8$. In this case, even with $\lambda=0.1$, the trend captures the data fluctuations more effectively than in the case without regularization, and the low-frequency component of the AR term is eliminated. The results for $\lambda=1$ are nearly identical to the case of $\lambda=0.1$. When $\lambda=16$, the trend becomes more flexible, and the variability of the AR component is further reduced. Compared to the case of $m_1=1$, the estimates appear to be improved.

Figure \ref{Fig_L2-estimates_M1=2} presents the results for the case of a second-order trend ($m_1=2$) with the AR model order set to $m_3=8$. 
Even with a small regularization parameter ($\lambda=0.1$), the trend captures the fluctuations in the data more effectively than in the unregularized case ($\lambda=0$), and the low-frequency component of the AR term is largely eliminated. The results for $\lambda=1$ are nearly identical to those for $\lambda=0.1$, indicating stability in the decomposition. When $\lambda=16$, the trend becomes more flexible, and the AR component exhibits further reduced variability. Compared to the case of $m_1=1$, the estimates under $m_1=2$ appear to provide a more robust and interpretable decomposition.

\newpage
\subsection{AR Modeling with $L_1$ Regularization Term}
In this subsection, we consider $L_1$ regularization method and estimate the parameter by minimizing
\begin{eqnarray}
  \log\ell (\theta) + \lambda R_1(\theta),
\end{eqnarray}
where regularization term $R_1(\theta)$ is defined by
\begin{eqnarray}
   R_1(\theta) &=& \sum_{j=1}^{m_3} |\theta_{n_v+j}| .  \nonumber 
\end{eqnarray}

\begin{figure}[htbp]
\begin{center}
\includegraphics[width=140mm,angle=0,clip=]{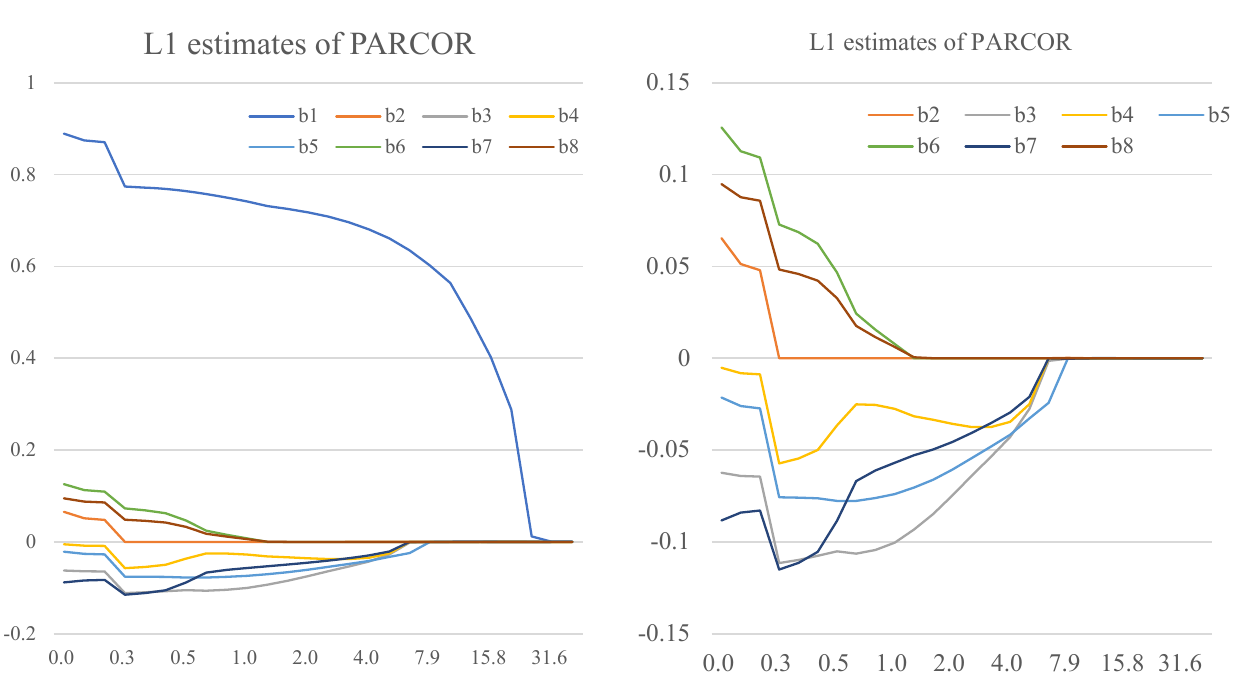}
\end{center}
\caption{Change of estimated parameters under $L_1$ regularization. $\lambda = 0$ and $10^{j/10},
j=-8,\ldots ,16$ }\label{Fig_L1-estimates}
\end{figure}

The left panel of Figure \ref{Fig_L1-estimates} illustrates how the partial autocorrelation coefficients (PARCOR) vary with the regularization parameter $\lambda$ for the Blsallfood data under the second-order trend model ($m_1=2$). 
The horizontal axis represents the value of $\lambda$, while the vertical axis shows the PARCOR values. As is well known, unlike $L_2$ regularization, $L_1$ regularization tends to drive an increasing number of PARCOR coefficients exactly to zero as $\lambda$ increases. A particularly notable feature, in contrast to $L_2$ regularization, is the relatively slower decay of the first PARCOR coefficient $b_1$, which retains a large absolute value even as other coefficients diminish.

\begin{figure}[tbp]
\begin{center}
\includegraphics[width=140mm,angle=0,clip=]{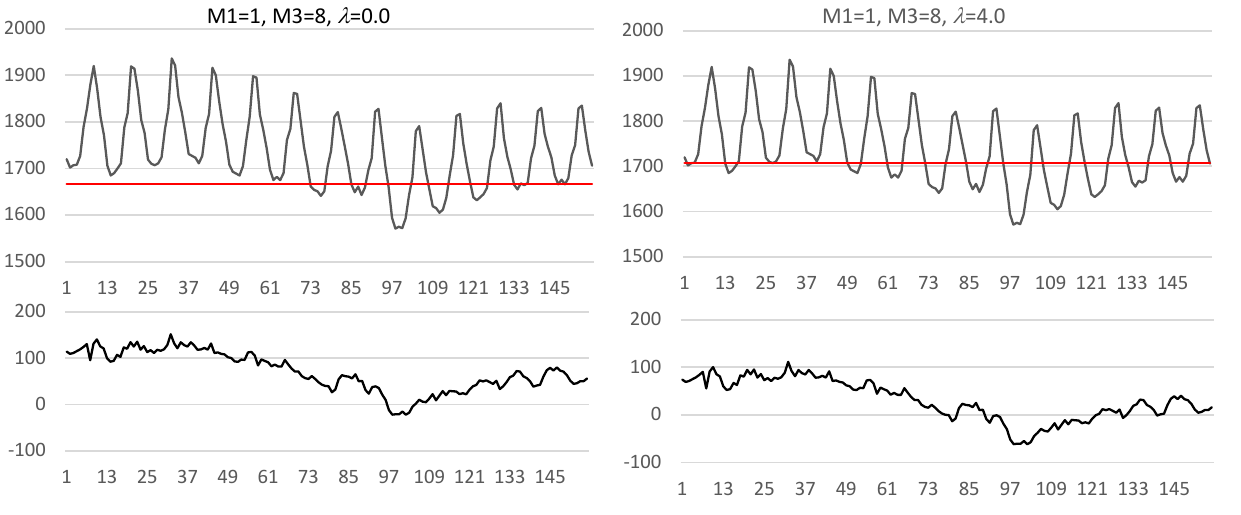}
\includegraphics[width=140mm,angle=0,clip=]{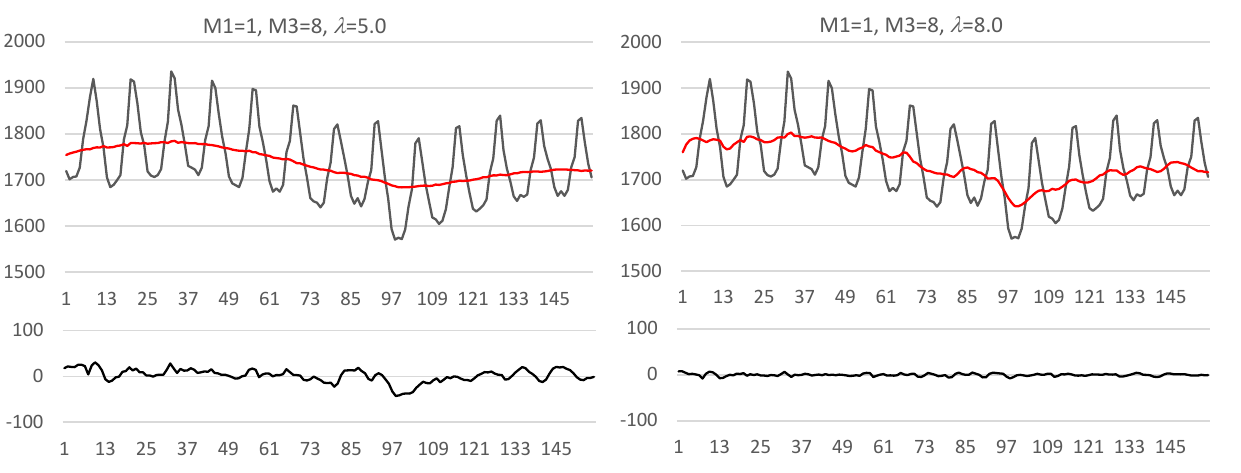}
\end{center}
\caption{Decomposition of time series by first order trend under various $L_1$ regularization terms. $\lambda = 0$ 4, 5 and 8 }\label{Fig_L1-estimates_m1=1}
\end{figure}

Figure \ref{Fig_L1-estimates_m1=1} presents the trend and AR components for the case of $m_1=1$ under four different values of the regularization parameter: $\lambda$=0, 4, 5, and 8. When $\lambda=4$, the trend remains nearly constant, similar to the case without regularization ($\lambda=0$), and the AR component retains long-period fluctuations. In contrast, at $\lambda=5$, the trend begins to reflect the overall pattern of the data more effectively, with the AR component showing a reduction in low-frequency variability. As $\lambda$ increases to 8, the trend exhibits greater variability, while the AR component becomes further diminished. It is important to note that with $L_1$ regularization, even small changes in $\lambda$ can result in substantial alterations to the estimated trend and AR components, highlighting the model's sensitivity to the regularization parameter.

\begin{figure}[tbp]
\begin{center}
\includegraphics[width=140mm,angle=0,clip=]{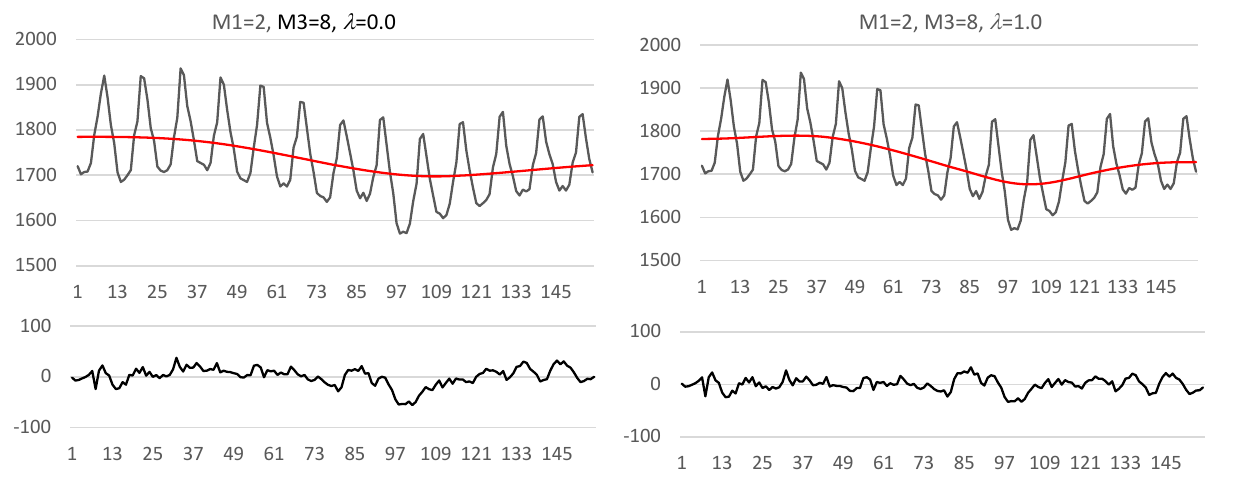}
\end{center}
\caption{Decomposition of time series by second order trend under various $L_1$ regularization terms. $\lambda = 0$ and 1 }\label{Fig_L1-estimates_m1=2}
\end{figure}

Figure \ref{Fig_L1-estimates_m1=2} shows the results for the case where $m_1 = 2$ and the AR model order is set to $m_3 = 8$. As illustrated in the left panel, even without a regularization term, the estimated trend partially captures the overall pattern of the data; however, the AR component exhibits long-period fluctuations. When the regularization parameter is set to $\lambda = 1$, the trend more accurately reflects the underlying tendency of the data, and the long-period fluctuations in the AR component appear to have been suppressed.

\section{Concluding Remarks}

In this paper, seasonal adjustment methods incorporating autoregressive (AR) component models are examined using empirical data, leading to the following key findings:

\begin{itemize} 
\item Within the framework of state-space modeling, when an AR component is included, models with and without observation noise tend to produce nearly identical results. Therefore, for seasonal adjustment procedures that incorporate AR components, it is methodologically sufficient to consider models without observation noise.

\item In seasonal adjustment models that include AR components, the trend component is often excessively smoothed, with long-period variations being misallocated to the AR component. This issue is particularly evident in first-order trend models, where the trend may degenerate into a constant. Nonetheless, for the empirical datasets considered in this study, it was found that imposing constraints on the modulus and argument of the AR model's eigenvalues yields more appropriate trend estimates.

\item In both $L_1$ and $L_2$ regularization, increasing the regularization parameter $\lambda$ causes the estimated trend to shift from a smooth curve to one that better captures the fluctuation characteristics of the time series. However, automatic selection of $\lambda$ remains a challenging task. Under $L_1$ regularization, the estimated coefficients exhibit high sensitivity to changes in $\lambda$. While PARCOR coefficients derived from $L_2$ regularization tend to gradually decrease in magnitude as $\lambda$ increases, those under $L_1$ regularization increasingly shrink to zero, resulting in a sparser representation. 
\end{itemize}



\begin{thebibliography}{2}

\item Akaike, H. (1980b), ^^ ^^ Seasonal adjustment by a Bayesian modeling", {\it J. Time Series Anal.}, {\bf 1}, 1--13.

\item Akaike, H. and Ishiguro, M. (1983), \lq\lq Comparative study of X-11 and Bayesian procedure of seasonal adjustment," {\it Applied Time Series Analysis of Economic Data}, U.S. Census Bureau.

\item Box, G.E.P., Hillmer, S.C. and Tiao, G.C. (1978), ^^ ^^ Analysis and modeling of seasonal time series", in {\it Seasonal Analysis of Time Seres}, ed.Zellner, A., US Bureau of the Census, {\it Economic Research Report ER-1}, 309--334.

\item Cleveland W. S. and Tiao G. C. (1976), ^^ ^^ Decomposition of seasonal time series: a model for the Census X-11 program", \textit{J. Am. Statist. Assoc.}, \textbf{11}, 581--587.

\item Findley, D. F., Monsell, B. C., Bell, W. R., Otto, M. C., and Chen, B. C. (1998). ^^ ^^ New capabilities and methods of the X-12-ARIMA seasonal-adjustment program,"  \textit{Journal of Business \& Economic Statistics}, \textbf{16}(2), 127--152.

\item Gersch, W., and Kitagawa, G. (1983), ^^ ^^ The prediction of time series with trends and seasonalities,"  \textit{Journal of Business \& Economic Statistics}, \textbf{1}(3), 253--264.


\item Hillmer S. C. and Tiao G. C. (1982), ^^ ^^ An ARIMA based approach to seasonal adjustment,"  \textit{J. Am. statist. Assoc.}, \textbf{11}, 63--70.

\bibitem{Kitagawa 1987}
Kitagawa, G. (1987). ^^ ^^ Non-Gaussian state-space modeling of nonstationary time series,"
{\it Journal of the American Statistical Association}, 82(400), 1032--1041.

\bibitem{Kitagawa 1989}
Kitagawa, G. (1989). ^^ ^^ Non-Gaussian seasonal adjustment," {\it Computers \& Mathematics with Applications}, 
Vol.18, No.6/7, 503--514.




\bibitem{Kitagawa 1996} 
Kitagawa, G. (1996). ^^ ^^ Monte Carlo filter and smoother for non-Gaussian nonlinear state space models," 
{\it Journal of Computational and Graphical Statistics}, \textbf{5}(1), 1--25.

\bibitem{Kitagawa 2021}
Kitagawa, G. (2021). \textit{Introduction to Time Series Modeling with Applications in R},
Second Edition, Chapman \& Hall CRC Press.

\item Kitagawa, G. and Gersch, W. (1984), ^^ ^^ A smoothness priors-state space modeling of
time series with trend and seasonality," {\it J. Amer. Statist. Assoc.}, {\bf 79}, 378--389.

\item Kitagawa, G. and Gersch, W. (1996), {\it Smoothness Priors Analysis of Time Series},
{\it Lecture Notes in Statistics}, {\bf 116}, Springer, New York.


\end{thebibliography}
\end{document}